\documentclass[twocolumn]{revtex4}
\usepackage{graphicx}
\usepackage{bm}
\usepackage{color}
\usepackage{amsmath}
\usepackage{natbib}

\begin{document}

\title{Nonlinear separate spin evolution in degenerate electron-positron-ion plasmas}

\author{Z. Iqbal}%
\email{abbasiravian@yahoo.com}
\affiliation{Department of Physics, G. C. University Lahore, Katchery Road, Lahore 54000, Pakistan
and Faculty of physics, Lomonosov Moscow State University, Moscow, Russian Federation.}

\author{Pavel A. Andreev}
\email{andreevpa@physics.msu.ru}
\affiliation{Faculty of physics, Lomonosov Moscow State University, Moscow, Russian Federation.}

\date{\today}

\begin{abstract}
The non-linear evolution of
spin-electron acoustic, positron-acoustic,
and spin-electron-positron acoustic waves is considered.
It is demonstrated that weakly nonlinear dynamics of each
wave leads to the soliton formation. Altogether, we report on existence of three different solitons. The spin-electron acoustic soliton known for electron-ion plasmas is described for electron-positron-ion plasmas for the first time. The existence of the spin-electron-positron acoustic soliton is reported for the first time. The positron-acoustic soliton and the spin-electron-positron acoustic soliton arise as the areas of a positive electric potential. The spin-electron acoustic soliton behaves as the area of a negative electric potential at the relatively small positron imbalance $n_{0p}/n_{0e}=0.1$ and as the area of a positive electric potential at the relatively large positron imbalance $n_{0p}/n_{0e}=0.5$.
\end{abstract}


\maketitle

\section{Introduction}

Electron-positron (e-p) plasmas exists in some astrophysical objects such as active galactic nuclei
\cite{Miller-1978}, in neutron star magnetosphere \cite{Paul-astro-phys-2007, sturrock-Astrophys-J-164-1971}.
Naturally, the existence
of electron-positron plasma in compact stars has been investigated and
number density ($1.1 \times 10^{33}cm^{-3}$) of such pair plasmas was
calculated by applying a simplified model of a gravitationally collapsing or
pulsating baryon core \cite{Han-Phys-Rev-D 86-2012}.
It is well-known
that the wave propagation phenomenon in e-p plasma is
different as in ordinary electron-ion (e-i) plasma. For example, it is
reported that many wave phenomena like acoustic waves, whistler waves,
Faraday rotation, lower hybrid waves and shear Alfven waves are absent in
the nonrelativistic e-p plasmas \cite{Iwamoto-Rev-E-47,
zank-Phys-Rev-E-51-1995}. This behavior of pair plasma is due to its
mass symmetry and charge anti-symmetry.

In order to study the dynamics of pair plasma in a controlled laboratory
environment, it is proposed that the interaction of ultrashort laser
pulses with gaseous or solid targets could lead to the generation of the
optically thin e-p plasma with above solid state densities in the range of
($10^{23}-10^{28}$) cm$^{-3}$ \cite{Wang -phys-plsma-20-2013}.
The degenerate e-p
plasma with ions is believed to be found in compact astrophysical objects
like neutron stars and the inner layers of white dwarfs \cite{Lai-Rev. Mod.
Phys. 73-2001, Shapiro-Germany-2004}.

The ion-acoustic wave (IAW) is caused by the density perturbation and is
amongst the most well studied electrostatic modes in both linear and non
linear regimes \cite{Davidson 1972, Nicolson 1983, Hasegawa} in e-i
plasmas.
The solitary structure of IAWs is studied in an unmagnetized electron-positron-ion (e-p-i) plasma \cite{Pope-1975}.
The nonlinear analysis shows that the amplitude of the
electron density hump reduces due to the presence of positrons in the e-i
plasma.
Oblique propagation of nonlinear
IAW is studied in magnetized e-p-i plasma \cite{Mahmood-new-j plasma-2003}.
It is found that the amplitude of the solitary
structure changes with the imbalance of positrons and the speed
of the obliquely propagating soliton in a magnetized plasma become subsonic,
while it is supersonic in the unmagnetized plasma \cite{Mahmood-new-j
plasma-2003}. Using the QHD model the effects of Bohm potential on the
linear and nonlinear properties of the IAW in an unmagnetized electron-ion
plasma is considered in \cite{Haas-2003-pop}. It is found that the
quantum effects modify the linear wave frequency and the variation of these
quantum effects are responsible to produce the shock waves, the bright and
the dark solitons of the IAWs.

Linear analysis of positron-acoustic waves (PAW) in e-p-i spinless quantum
plasma is discussed in \cite{Azhar-2011, Tsintsadze-j.plasma phy-2013}
and non linear properties of PAW is presented in \cite{Nejoh--Aust-
J-Phys, Metref-pop-2014, Tribeche-pop-2009}. For instance, Nejoh \cite{Nejoh--Aust- J-Phys}
is studied the nonlinear propagation of PAWs in an e-p plasma with an electron beam.

Recently, the subject of spin-1/2 quantum plasma attracts the attention of
plasma physicist due to the existence of new wave phenomena
\cite{Vagin-2006, Moscow University Physics Bulletin-2007, Browdn-PRL 101-
(2008), Misra-j-plasma phys-2010, Hussain PP 14 spin bernst, Andreev PRE 15,
Pavel-Zafar-2016}. The subject of spin-1/2 quantum plasma was first explored
by Kuzmenkov et.al. in 2001 \cite{kuzmenkov- page-136-2001, kuzmenkov- page-258-2001}
by presenting the method of derivation of many particle quantum
hydrodynamic (QHD) equations.

The generalization of method presented by Kuzmenkov et.al. \cite{kuzmenkov-
page-136-2001, kuzmenkov- page-258-2001} is given in \cite{Andreev PRE 15,
Pavel-soliton-pop-2016}. In \cite{Andreev PRE 15} the author presents the
method of the separated spin evolution (SSE) QHDs for the degenerate electrons in
which the electrons with spin-up and spin-down treated as two different
fluids. It is essential if the populations of spin-up electrons $n_{u}$
and spin-down electrons $n_{d}$ in the presence of external magnetic field
is different $n_{u}\neq n_{d}$. This difference of populations of quantum
states is responsible for difference of Fermi pressures of the spin-up and
spin-down electrons. It lead to a new type of a soundlike solution
called the spin-electron acoustic wave (SEAW).

However, at the consideration of oblique propagation of longitudinal waves
in magnetized plasma a pair of SEAWs is reported in \cite{Pavel- annal
phys- 15}. For the wide-ranging studies of SEAWs the SSE quantum kinetics is derived in \cite{PA-arXiv-1409-2014
Landau damping}. Analyzing the Landau damping of the SEAWs, it is
concluded that these waves are weakly damped. With the aid of SSE-QHD the existence of SEAW
and its properties in two dimensional
electron gas situated in an external magnetic field is described in
\cite{Pavel-EPL-113 -2016}. It is demonstrated that the Langmuir wave is
replaced by two hybrid waves one of them is modified Langmuir wave and other
is SEAW. Further, the SSE-QHD equations were applied to find the existence of
surface spin-electron acoustic wave in magnetically ordered metals in \cite{Pavel-surface -2015}.
It was shown that in the regime of small
polarization the hybridization occurred between the surface Langmuir wave
and the surface SEAW. Another important application of the SSE-QHD method is
the electron-spelnon (spelnon is the quanta of SEAW) interaction
which is responsible for the Cooper pair formation giving a mechanism for the
high-temperature superconductivity \cite{P. A-superconductivity- Oct-2015}.
The SSE-QHD method is further generalized by including exchange
interaction effects \cite{Pavel-soliton-pop-2016}. This generalization is
applied to study the solitary structure of SEAW in the degenerate e-i plasma.

Applications of the SSE-QHD
in e-p-i plasma leads to the existence of
four longitudinal waves either the propagation parallel or perpendicular to an
external magnetic field \cite{Pavel-Zafar-2016}: Langmuir, SEAW, PAW, spin electron-positron acoustic
waves (SEPAW). At the oblique propagation the spectrum consists of eight longitudinal
waves \cite{Pavel-Zafar-2016}: the Langmuir wave,
the Trivelpiece--Gould wave, the pair of PAWs, the pair of SEAWs, and the pair of SEPAWs.
In the present work, we discuss the effects
of spin polarization and variation of concentration ratio $n_{p}/n_{e}$ on
the solitary structures of PAW and recently investigated SEAW and SEPAWs.

This paper is organized as follows.
In Sec. II we describe the basic equations, formulate problem of weakly non-linear evolution, and consider the first order approximation.
In Sec. III we present soliton solutions arising in the second order approximation.
In Sec. IV we present numerical analysis for three kinds of soliton solutions associated with the SEAW, PAW, and SEPAW.
In Sec. V a brief summary of obtained results is presented.

\section{Model}

For the description of the electron-positron-ion plasma, we apply the SSE-QHD
equations which are recently derived in \cite{Andreev PRE 15, Pavel- annal
phys- 15}. The SSE is considered at the action of the external magnetic
field. In this case, each spin-1/2 species containing spin-up and spin-down particles is considered as two different
fluids. In this paper, we describe the solitary structure of the recently
described longitudinal (SEAW, PAW and SEPAW) waves.

Continuity equations for each spin projection of each species arise as \cite{Andreev PRE 15}
\begin{equation}\label{SEASolEPI_cont eq}
\partial_{t}n_{as}+\nabla (n_{as}\mathbf{v}_{as})=(-1)^{i_{s}}T_{az},
\end{equation}
where $a=e,p$ for electrons and positrons correspondingly, $s=u,d$ for the
spin-up and spin-down states of particles, $n_{as}$ and $\mathbf{v}_{as}$
are the concentration and velocity field of particles of species $a$ being
in the spin state $s$, $T_{az}=\frac{\gamma_{a}}{\hbar}(B_{x}S_{ay}-B_{y}S_{ax})$ is the z-projection of spin torque,
$i_{s}$: $i_{u}=2$, $i_{d}=1$, $\gamma_{a}$ is the magnetic moment, $\mid\gamma_{e}\mid=\gamma_{p}\approx\mu_{B}$, $\mu_{B}$ is the Bohr magneton, with the spin density projections $S_{ax}$ and $S_{ay}$,
each of them simultaneously describe evolution of the spin-up and spin-down particles of each species.
Therefore, functions $S_{ax}$ and $S_{ay}$ do not have subindexes $u$ and $d$. In this model, the z-projection
of the spin density $S_{az}$ is not an independent variable. It is a
combination of concentrations $S_{az}=n_{au}-n_{ad}$. The time evolution of
the velocity fields of each species of particles for each projection of spin $v_{au}$ and $v_{ad}$ is governed by the Euler equations \cite{Andreev PRE 15}
$$mn_{as}(\partial _{t}+\mathbf{v}_{as}\nabla )\mathbf{v}_{as}+\nabla P_{as}$$
$$=q_{a}n_{as}\left(\mathbf{E}+\frac{1}{c}[\mathbf{v}_{as},\mathbf{B
}]\right) +(-1)^{i_{s}}\gamma _{a}n_{as}\nabla B_{z} $$
\begin{equation}\label{SEASolEPI Euler eq}+\frac{\gamma_{a}}{2}(S_{ax}\nabla B_{x}+S_{ay}\nabla
B_{y})+(-1)^{i_{s}}m(\widetilde{\mathbf{T}}_{az}-\mathbf{v}_{as}T_{az}), \end{equation}
with $P_{as}=(6\pi^{2})^{\frac{2}{3}}n_{as}^{\frac{5}{3}}\hbar^{2}/5m$, $
\widetilde{\mathbf{T}}_{az}=\frac{\gamma_{a}}{\hbar}(\mathbf{J}_{(M)ax}B_{y}-\mathbf{J}_{(M)ay}B_{x})$, which is the torque current, where $\mathbf{J}_{(M)ax}=(\mathbf{v}_{au}+\mathbf{v}_{ad})S_{ax}/2$,
and $\mathbf{J}_{(M)ay}=(\mathbf{v}_{au}+\mathbf{v}_{ad})S_{ay}/2$ are the convective
parts of the spin current tensor. $\textbf{E}$ and $\textbf{B}$ are the electric
and magnetic fields.

The Coulomb exchange interaction affects the dynamics of electrons and positrons at concentrations below $10^{25}$ cm$^{-3}$ \cite{Pavel-soliton-pop-2016, Andreev AoP 14 exchange}. We consider the concentrations above $10^{27}$ cm$^{-3}$. Hence, we do not consider the exchange interaction in hydrodynamic equations (\ref{SEASolEPI_cont eq}), (\ref{SEASolEPI Euler eq}).

In equilibrium, we assume that the stationary positive ions
are present in background. So, the neutrality condition is $n_{eu}+n_{ed}=n_{pu}+n_{pd}+n_{i}$,
where $n_{i}$ is the concentration of ions. For the propagation of longitudinal waves in
electron-positron-ion plasma the Poisson equation can be used
\begin{equation} \label{SEASolEPI div E}
\nabla\cdot\textbf{E}=4\pi e(n_{i}+n_{pu}+n_{pd}-n_{eu}-n_{ed})
\end{equation}
Since we consider the propagation of longitudinal waves parallel to the external
magnetic in e-p-i plasma. Hence, the perturbation of magnetic field is
zero. Equations (\ref{SEASolEPI_cont eq})-(\ref{SEASolEPI Euler eq}) can be written in the following
simplified form
\begin{equation} \label{SEASolEPI Euler short}
mn_{as}(\partial_{t}+\mathbf{v}_{as}\nabla)\mathbf{v}_{as} +\nabla P_{as}=-q_{a}n_{as}\nabla \phi,
\end{equation}
\begin{equation} \label{SEASolEPI cont short}
\partial _{t}n_{as}+\nabla (n_{as}\mathbf{v}_{as})=0,
\end{equation}
\begin{equation} \label{SEASolEPI lapl of phi with short eq}
\nabla^{2}\phi =4\pi e(n_{eu}+n_{ed}-n_{pu}-n_{pd}-n_{i}),
\end{equation}
where we have used $\textbf{E}=-\nabla\phi$ and $\phi$ is the
potential of electric field. To study the nonlinear dynamics of the spin-1/2
quantum magnetized e-p-i plasma in the SSE, we employ the reductive
perturbation method \cite{Washimi-1966} and transform space and time coordinates
to the stretching coordinates $\xi=\epsilon^{\frac{1}{2}}(z-Vt)$
and $\tau=\epsilon^{\frac{3}{2}}t$, where $V$
is the wave phase speed and $\epsilon$ is a small dimensionless parameter
measuring the amplitude of the perturbation. The hydrodynamic variables are
expanded about their equilibrium values in powers $\epsilon$ of as
\begin{equation} \label{SEASolEPI phi expansion}
\phi=\epsilon \phi_{1}+\epsilon^{3}\phi_{2},
\end{equation}
\begin{equation} \label{SEASolEPI n expansion}
n_{as}=n_{0as}+\epsilon n_{1as}+\epsilon^{2}n_{2as},
\end{equation}
\begin{equation} \label{SEASolEPI v expansion}
\mathbf{v}_{as}=\epsilon \textbf{v}_{1as}+\epsilon^{2}\textbf{v}_{2as}.
\end{equation}
We substitute equations (\ref{SEASolEPI phi expansion})-(\ref{SEASolEPI v expansion})
in equations (\ref{SEASolEPI Euler short})-(\ref{SEASolEPI lapl of phi with short eq}).
The terms in the first order on $\epsilon$ from
the continuity equation and Euler equation give the expression for the
perturbed number density in term of the electric field as
\begin{equation} \label{SEASolEPI n 1 expression}
n_{1as}=-\sum_{a=e,p}\left(\frac{n_{0us}\frac{q_{a}}{m}\phi_{1}}{\left(U_{Fau}^{2}-V^{2}\right)} +\frac{n_{0ad}\frac{q_{a}}{m}\phi_{1}}{\left(U_{Fad}^{2}-V^{2}\right)}\right),
\end{equation}
where $U_{Fas}^{2}=\frac{1}{3}\frac{\hbar^{2}}{m^{2}}\left(6\pi^{2}n_{as}\right) ^{\frac{2}{3}}$. Using the Eq. (\ref{SEASolEPI n 1 expression}) in the Poisson
equation in the first order on $\epsilon$ which is given as $n_{1eu}+n_{1ed}-n_{1pu}-n_{1pd}=0,$ we find
$$\frac{1-\eta_{e}}{\left(1-\eta_{e}\right)^{\frac{2}{3}}-w^2}
+\frac{\beta\left(1-\beta^{-\frac{2}{3}}\eta_{e}\right)}{\beta^{\frac{2}{3}}\left(1-\beta^{-\frac{2}{3}}\eta_{e}\right)^{\frac{2}{3}}-w^2}$$
\begin{equation}\label{SEASolEPI eq for V}
+     \frac{1+\eta_{e}}{\left(1+\eta_{e}\right)^{\frac{2}{3}}-w^2}
+\frac{\beta\left(1+\beta^{-\frac{2}{3}}\eta_{e}\right)}{\beta^{\frac{2}{3}}\left(1+\beta^{-\frac{2}{3}}\eta_{e}\right)^{\frac{2}{3}}-w^2}=0 \end{equation}
in accordance with $n_{0eu}=\frac{1}{2}n_{0}(1-\eta_{e})$, $n_{0ed}=\frac{1}{2}n_{0}(1+\eta_{e})$, $n_{0pu}=\frac{1}{2}n_{0}\beta\left(1+\eta_{p}\right)$, $n_{0pd}=\frac{1}{2}n_{0}\beta\left(1-\eta_{p}\right)$ and $\eta_{p}=\beta^{-\frac{2}{3}}\eta_{e}$, where the imbalance of positrons $\beta =n_{p}/n_{e}$ and the spin-polarization $\eta_{e}=3\mu_{B} B_{0}/2\varepsilon_{Fe}$, with the Fermi energy $\varepsilon_{Fe}=mv_{Fe}^{2}/2$ and $v_{Fe}=\frac{\hbar}{m}\left(3\pi^{2}n_{0e}\right)^{\frac{1}{3}}$ is Fermi velocity.

\begin{figure}
\includegraphics[width=8cm,angle=0]{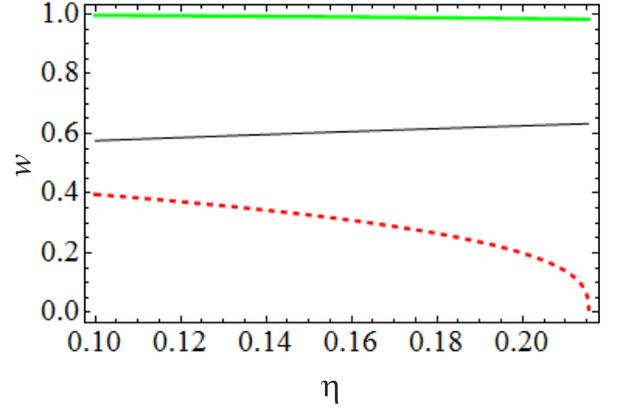}
\caption{\label{SEASolEPI F V as F of h} (Color online) The figure shows the velocity dependence on the spin polarization $\eta$ at $\beta=0.1$, where $\eta$ is the spin polarization of electrons while polarization of positrons is represented via $\eta$.}
\end{figure}

\begin{figure}
\includegraphics[width=8cm,angle=0]{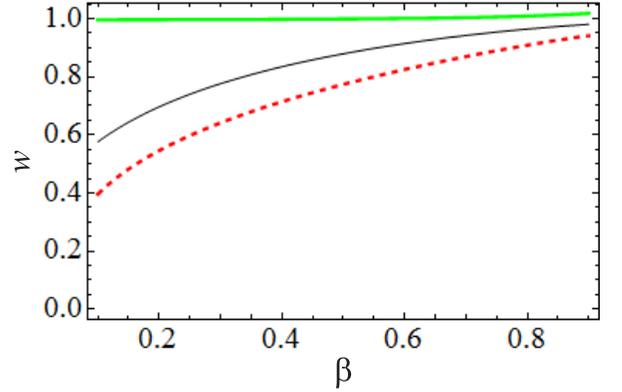}
\caption{\label{SEASolEPI F V as F of b} (Color online) The figure shows the velocity dependence on the imbalance of positrons $\beta=n_{0p}/n_{0e}$ at a fixed spin polarization $\eta=0.1$.}
\end{figure}

Equation (\ref{SEASolEPI eq for V}) is a cubic equation on dimensionless phase velocity $w^{2}$ defined as $V^{2}=w^{2}v_{Fe}^{2}/3$.
Each solution of above equation corresponds to the linear solution of each
longitudinal wave among the three waves found in Ref. \cite{Pavel-Zafar-2016} i.e. SEPAW, PAW and SEAW. Dependencies of these velocities on the spin-polarization $\eta_{e}$ and the positron imbalance $\beta$ are given in Figs. \ref{SEASolEPI F V as F of h} and \ref{SEASolEPI F V as F of b}.


\section{Solitons}

Further application of the reductive perturbation method \cite{Washimi-1966} in the second order on $\varepsilon$ leads to the nonlinear
Korteweg–de Vries (KdV) equation in terms electrostatic potential
\begin{equation} \label{SEASolEPI KdV eq}
D\partial_{\tau}\phi_{1}-R\partial_{\xi}\phi_{1}^{2}+\partial_{\xi}^{3}\phi _{1}=0.
\end{equation}
Coefficients $D$ and $R$ are given below.

In order to find the solitary wave solution, we introduce $\sigma=\zeta-U_{0}\tau$, where $U_{0}$ is
the soliton propagation velocity. Next, we integrate the KdV equation (\ref{SEASolEPI KdV eq}) and apply the boundary conditions such that $\phi_{1}$ and its derivatives
approaches to zero at $\sigma \rightarrow \pm\infty$. This yields the
well-known soliton solution of KdV equation (\ref{SEASolEPI KdV eq})
\begin{equation}\label{SEASolEPI sol for phi}
\phi _{1}=-\frac{3DU_{0}}{2R}\frac{1}{\cosh^{2}[\frac{\sqrt{DU_{0}}\sigma}{2}]}.
\end{equation}

The explicit form of the coefficients $D$ and $R$ are found to be as follows
\begin{equation}\label{SEASolEPI D gen str} D=2V\sum_{s=u,d}\left( \frac{\omega_{Les}^{2}}{(U_{Fes}^{2}-V^{2})^{2}}+
\frac{\omega_{Lps}^{2}}{(U_{Fps}^{2}-V^{2})^{2}}\right),\end{equation}
and
\begin{widetext}
\begin{equation}\label{SEASolEPI R gen str} R=\frac{e}{m}\sum_{s=u,d}\left[\frac{\omega_{Les}^{2}}{(U_{Fes}^{2}-V^{2})^{2}} \left(\frac{1}{2}+\frac{V^{2}+\frac{1}{3}U_{Fes}^{2}}{V^{2}-U_{Fes}^{2}}\right)
-\frac{\omega_{Lps}^{2}}{(U_{Fps}^{2}-V^{2})^{2}}
\left(\frac{1}{2}+\frac{V^{2}+\frac{1}{3}U_{Fps}^{2}}{V^{2}-U_{Fps}^{2}}
\right) \right].\end{equation}
From the Eq. (\ref{SEASolEPI sol for phi}) it is clear that the properties of solitons depend upon the
signs of coefficients $D$ and $R$. For the numerical analysis of the soliton solution
(\ref{SEASolEPI sol for phi}) we write the coefficients $D=\sqrt{3}D_{0}/(v_{Fe}r_{De}^{2})$ and $R=3eR_{0}/(2m v_{Fe}^{2}r_{De}^{2})$ in term of dimensionless parameters:
\begin{equation}\label{SEASolEPI D 0}
D_{0}=
\left[\frac{w (1-\eta_{e})}{\left(\left(1-\eta_{e}\right)^{\frac{2}{3}}-w^2 \right)^{2}}
+     \frac{w (1+\eta_{e})}{\left(\left(1+\eta_{e}\right)^{\frac{2}{3}}-w^2 \right)^{2}}\right.
+\frac{w\beta
\left(1-\beta^{-\frac{2}{3}}\eta_{e}\right)}{\left(\beta^{\frac{2}{3}}\left(1-\beta^{-\frac{2}{3}}\eta_{e}\right)^{\frac{2}{3}}-w^2 \right)^{2}}\left.
+\frac{w\beta
\left(1+\beta^{-\frac{2}{3}}\eta_{e}\right)}{\left(\beta^{\frac{2}{3}}\left(1+\beta^{-\frac{2}{3}}\eta_{e}\right)^{\frac{2}{3}}-w^2 \right)^{2}}\right],
\end{equation}
and
$$R_{0}=
\left[\frac{(1-\eta_{e})}{\left(\left(1-\eta_{e}\right)^{\frac{2}{3}}-w^2 \right)^{2}}\left(\frac{1}{2}+\frac{w^2 +\frac{1}{3}\left(
1-\eta_{e}\right)^{\frac{2}{3}}}{\left(w^2 -\left(1-\eta_{e}\right)^{\frac{2}{3}}\right)}\right)\right.
-\frac{\beta\left(1-\beta^{-\frac{2}{3}}\eta_{e}\right)}{\left(\beta^{\frac{2}{3}}\left(1-\beta^{-\frac{2}{3}}\eta_{e}\right)^{\frac{2}{3}}-w^2 \right)^{2}}\left(\frac{1}{2}+\frac{w^2 +\frac{1}{3}\beta^{\frac{2}{3}}\left(1-\beta^{-\frac{2}{3}}\eta_{e}\right)^{\frac{2}{3}}}{\left(w^2 -\beta^{\frac{2}{3}}\left(1-\beta^{-\frac{2}{3}}\eta_{e}\right)^{\frac{2}{3}}\right)}\right)$$
\begin{equation}\label{SEASolEPI R 0}
+\frac{(1+\eta_{e})}{\left(\left(1+\eta_{e}\right)^{\frac{2}{3}}-w^2 \right)^{2}}\left(\frac{1}{2}+\frac{w^2 +\frac{1}{3}\left(
1+\eta_{e}\right)^{\frac{2}{3}}}{\left(w^2 -\left(1+\eta_{e}\right)^{\frac{2}{3}}\right)}\right) -\frac{\beta\left(1+\beta^{-\frac{2}{3}}\eta_{e}\right)}{\left(\beta^{\frac{2}{3}}\left(1+\beta^{-\frac{2}{3}}\eta_{e}\right)^{\frac{2}{3}}-w^2 \right)^{2}}\left.\left(\frac{1}{2}+\frac{w^2 +\frac{1}{3}\beta^{\frac{2}{3}}\left(1+\beta^{-\frac{2}{3}}\eta_{e}\right)^{\frac{2}{3}}}{\left(w^2 -\beta^{\frac{2}{3}}\left(1+\beta^{-\frac{2}{3}}\eta_{e}\right)^{\frac{2}{3}}\right)}\right)\right]
\end{equation}\end{widetext}
where we used the dimensionless velocity of the soliton $w$ and the Debye radius $r_{De}^{2}=v_{Fe}^{2}/3\omega _{Le}^{2}$ of electrons.

Amplitudes of all solitons are proportional to $D/R=(2\sqrt{3}(3\pi^{2})^{1/3}\hbar/3e)n_{0e}^{1/3}D_{0}/R_{0}$. It shows that the amplitudes of all solitons increase with the increase of equilibrium concentration of electrons.

Coefficient $R$ can be equal to zero for the SEAW, as we show it below. In this case, amplitude of soliton $\phi_{1}$ goes to infinity. Our approximation of small perturbations does not work in the area of the divergence. Hence, we present our results on SEA soliton far from the divergence, in area of applicability of our approach.

As we have mentioned in the introduction the analysis of soliton solution of
SEAW for e-i plasma is presented in \cite{Pavel-soliton-pop-2016}.
In this paper, we discuss the soliton solutions of PAWs \cite{Nejoh--Aust- J-Phys, Metref-pop-2014, Tribeche-pop-2009} and recently found SEPAW
and SEAW waves in e-p-i plasma \cite{Pavel-Zafar-2016}. Although, the non-linear PAW has been presented for spinless plasma in literature
\cite{Nejoh--Aust- J-Phys, Metref-pop-2014, Tribeche-pop-2009}, but we first time
present the effect of spin polarization on the non-linear PAW.

\begin{figure}
\includegraphics[width=8cm,angle=0]{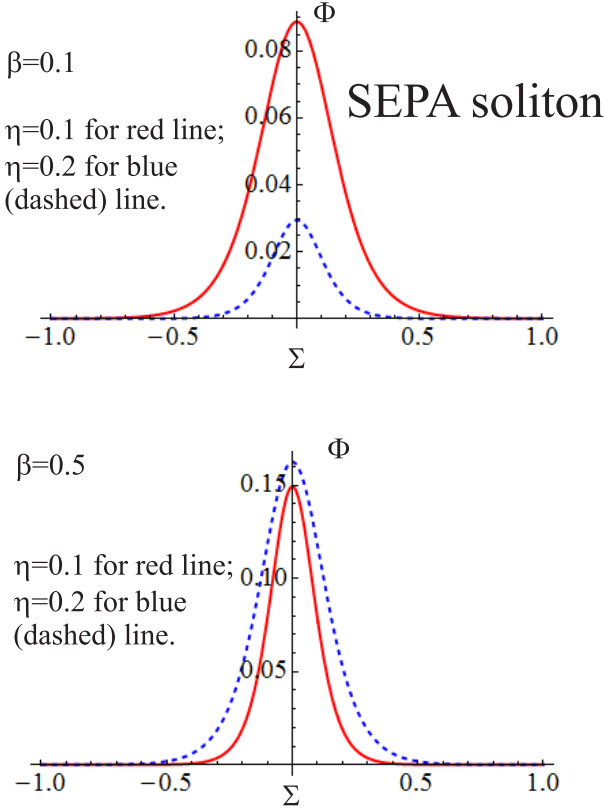}
\caption{\label{SEASolEPI F SEPAW b hh} (Color online) The figure shows the profile of the SEPA soliton at different parameters presented in the pictures. We keep fixed equilibrium concentration of electrons $n_{0e}=10^{27}$ cm$^{-3}$. Here and below, we change concentration of positrons to change $\beta$ which affects polarization of positrons. We change external magnetic field to change spin polarization of electrons and positrons with no effect on $\beta$. For $\eta_{e}=0.1$ we have $B_{0}=4.205\times 10^{10}$ G and for $\eta_{e}=0.2$ we have
$B_{0}=8.410\times 10^{10}$ G. Here and below, the amplitude of soliton $\phi_{1}$ and coordinate $\sigma$ are presented in a dimensionless forms $\Phi=\sqrt{3}e\phi_{1}/(2mv_{Fe}U_{0})$ and $\Sigma=\sqrt{\sqrt{3}U_{0}/v_{Fe}}\sigma/r_{De}$.}
\end{figure}

\begin{figure}
\includegraphics[width=8cm,angle=0]{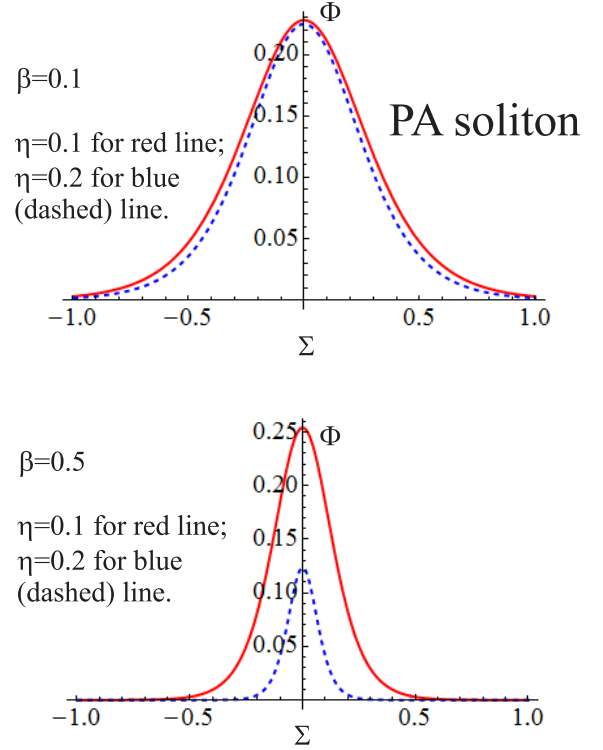}
\caption{\label{SEASolEPI F PAW b hh} (Color online) The figure shows the positron acoustic soliton with variations of the spin polarization $\eta_{e}$ and the positron imbalance $\beta$.}
\end{figure}

\begin{figure}
\includegraphics[width=8cm,angle=0]{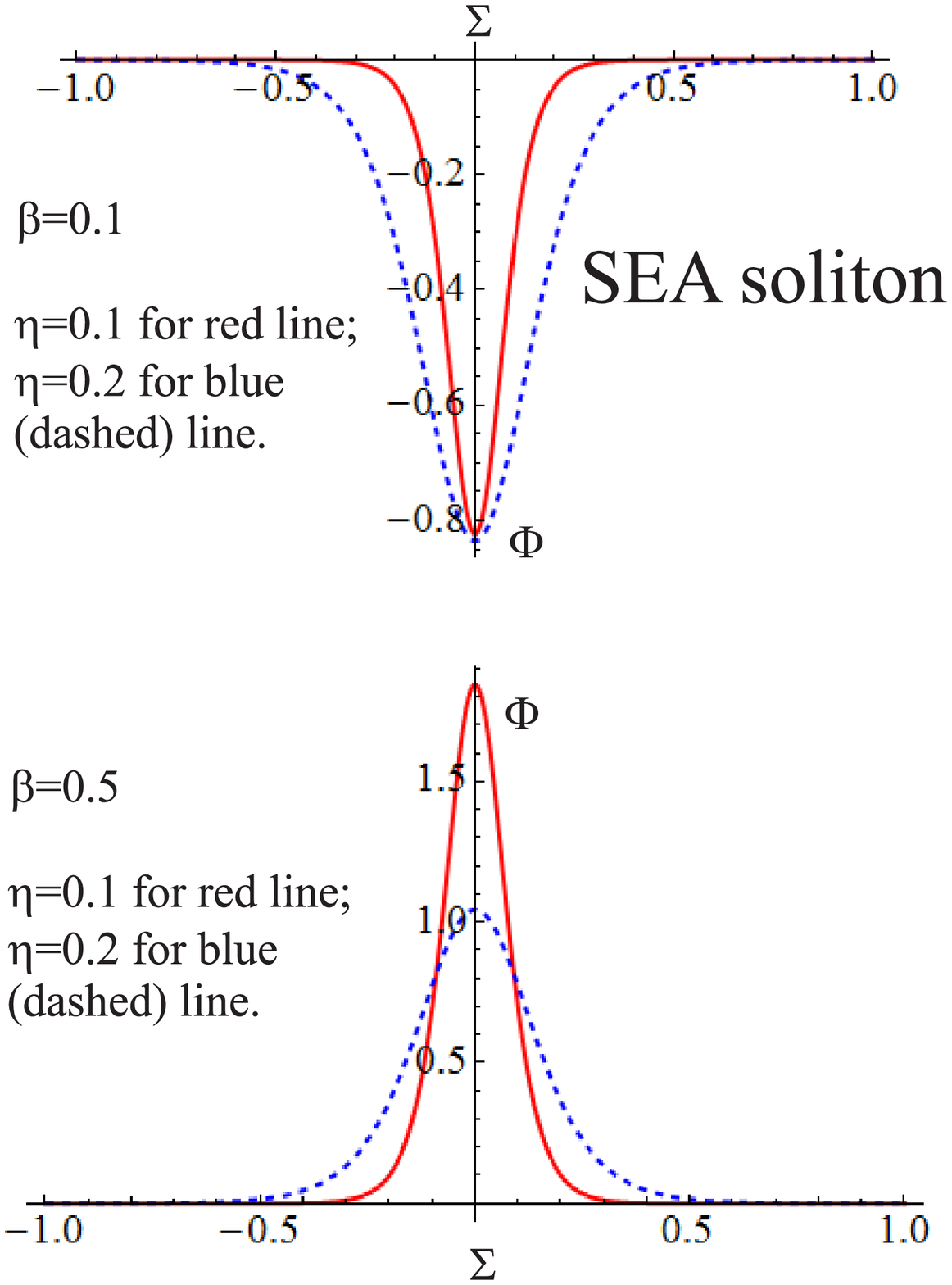}
\caption{\label{SEASolEPI F SEAW b hh} (Color online) The figure shows the effect of spin polarization $\eta_{e}$ and positron imbalance $\beta$ on the spin electron acoustic soliton.}
\end{figure}

\section{Numerical analysis of soliton solutions}

We work in area of parameters near the
equilibrium concentration $n_{0}=10^{27}$ cm$^{-3}$ and the external magnetic field $B_{0}=10^{10}$ G. Quasi stability of e-p-i plasmas at this parameters is follows from calculations presented in Refs. \cite{Svensson Astr J 82} and \cite{Akbari-Moghanjoughi PoP 10}.

The increase of the concentration $n_{e}$ leads to the decrease of the spin polarization since $\gamma_{e}=3\mu_{B} B_{0}/2\varepsilon_{Fe}$. Therefore, the area of larger concentrations corresponds to the area of smaller spin polarization.

Three branches of the phase velocity $V$ leads to formation of three different solitons. Fig. \ref{SEASolEPI F V as F of h} shows that the upper line describing SEAW slowly goes down showing the decrease of the phase velocity with the increase of the spin-polarization. The middle line describing PAW shows different behavior. Its phase velocity slowly increases with the increase of the spin-polarization. Its increase is faster in compare with the decrease rate of SEAW phase velocity. The lower line describes SEPAW. Its phase velocity quickly decreases at the increase of the spin-polarization. Fig. \ref{SEASolEPI F V as F of b} shows that all branches increases with the increase of the positron imbalance. At large positron imbalance difference of their phase velocity becomes rather small. In the following subsections, we describe each soliton.

\subsection{SEPA soliton}

As it was demonstrated in our previous paper \cite{Pavel-Zafar-2016}, the SEPAW existing in degenerate e-p-i plasmas is one of three waves with the linear spectrum at small $k$ which has minimal phase velocity. Hence, the SEPAW corresponds to the smallest solution in Figs. \ref{SEASolEPI F V as F of h} and \ref{SEASolEPI F V as F of b}. Substituting this solution in coefficients $D_{0}$ and $R_{0}$ presented by formulae (\ref{SEASolEPI D 0}) and (\ref{SEASolEPI R 0}), we obtain an explicit form of the SEPA soliton. This solution is presented in Fig. \ref{SEASolEPI F SEPAW b hh}. To the best of our knowledge, existence of the SEPA soliton is demonstrated in this paper for the first time. Let us describe properties of the SEPA soliton demonstrated in Fig. \ref{SEASolEPI F SEPAW b hh}.

Fig. \ref{SEASolEPI F SEPAW b hh} shows that at larger positron imbalance (the lower figure with $\beta=0.5$ in compare with the upper figure with $\beta=0.1$) the change of spin polarization make smaller effect. Moreover, the increase of spin polarization gives opposite effects at small $\beta=0.1$ and large $\beta=0.5$ positron imbalance.
At $\beta=0.1$ ($\beta=0.5$), the increase of spin polarization from $\eta=0.1$ to $\eta=0.2$ leads to the decrease (the increase) of the amplitude and width of SEPA soliton.

On the other hand, we can compare the soliton profiles at different $\beta$ at fixed spin polarization. We see that red (continuous) lines in the upper and the lower pictures in Fig. \ref{SEASolEPI F SEPAW b hh} corresponding to $\eta=0.1$ goes higher and thinner at the increase of $\beta$ from $\beta=0.1$ to $\beta=0.5$. It shows that amplitude increases and the width decreases.

At the larger polarization $\eta=0.2$ described by blue (dashed) lines in the upper and the lower pictures in Fig. \ref{SEASolEPI F SEPAW b hh} soliton becomes considerable higher and its width increases slightly at the increase of $\beta$ from $\beta=0.1$ to $\beta=0.5$.

In all considered regimes, the electric potential in the SEPA soliton is positive. Thus, we find that the SEPA soliton is a bright soliton of the electric potential.

\subsection{SEA soliton}

The bulk SEAW in e-i plasmas was predicted in Ref. \cite{Andreev PRE 15}. Corresponding soliton solution was found in Ref. \cite{Pavel-soliton-pop-2016}. Considering the SSE in e-p-i plasmas, in our previous paper, we described the properties of the linear SEAWs. We present now the behavior of SEA soliton in e-p-i plasmas.

In opposite to the SEPA soliton described in the previous subsection and the PA soliton described in the next subsection, the electric potential in the SEA soliton is negative at small positron imbalance $\beta=0.1$. In this regime, we obtain that the SEA soliton is a dark soliton of the electric potential.

Fig. \ref{SEASolEPI F SEAW b hh} shows that at the larger positron imbalance (the lower figure with $\beta=0.5$ in compare with the upper figure with $\beta=0.1$), the coefficient $R$ in the KdV equation (\ref{SEASolEPI KdV eq}) changes its sign. Hence, the electric potential in the SEA soliton is positive at the larger positron imbalance $\beta=0.5$ and the SEA soliton is a bright soliton of the electric potential, in this regime.

Fig. \ref{SEASolEPI F SEAW b hh} shows that at the larger positron imbalance the change of spin polarization make larger effect. Moreover, the increase of spin polarization gives opposite effects at small $\beta=0.1$ and large $\beta=0.5$ positron imbalance.
At $\beta=0.1$, the increase of spin polarization from $\eta=0.1$ to $\eta=0.2$ leads to the increase of the amplitude and width of SEPA soliton. The change of amplitude is rather small, while the width becomes almost two times larger.
At $\beta=0.1$, the increase of spin polarization from $\eta=0.1$ to $\eta=0.2$ decreases the module of amplitude making it approximately two times smaller. In this regime, the width increases becoming almost two times larger.

Considering a fixed spin polarization we can compare the soliton profiles at different $\beta$. As we have mentioned above, the change of $\beta$ leads to change of sign of the electric potential. This change happens due to change of $R$ located in the denominator of the amplitude. It means that solution tends to infinity in the area of $R=0$. The applied approximation requires small amplitude of the solutions. Thus, prediction near the area of $\beta$ and $\eta$ giving $R=0$ is not covered by our analysis.

We see that module of amplitude goes higher at the increase of $\beta$ from $\beta=0.1$ to $\beta=0.5$ in the following regime: $\eta=0.1$ described by the red (continuous) lines in the upper and the lower pictures in Fig. \ref{SEASolEPI F SEAW b hh}. The width does not show changes under described conditions.

At the larger spin polarization $\eta=0.2$ described by blue (dashed) lines in the upper and the lower pictures in Fig. \ref{SEASolEPI F SEAW b hh} module of the soliton amplitude slightly increases (on 20 percents). The width does not show visible changes.

\subsection{PA soliton}

Existence of the PAWs and their non-linear evolution was demonstrated in Ref. \cite{Nejoh--Aust- J-Phys}. Influence of the SSE and spin polarization on the spectrum of linear PAWs in considered in Ref. \cite{Pavel-Zafar-2016}. We describe now the non-linear evolution of PAWs at the account of SSE. The soliton profile at different values of parameters is presented in Fig. \ref{SEASolEPI F PAW b hh}.

The PA solitons arises as a bright soliton in terms of the electric potential.

The increase of spin polarization at fixed $\beta$ leads to the decrease of the positron-acoustic soliton amplitude and width Fig. \ref{SEASolEPI F PAW b hh}. This decrease is larger at the larger $\beta=0.5$ positron imbalance.

Next, we compare the soliton profiles at different $\beta$ at fixed spin polarization.
We see that red (continuous) lines in Fig. \ref{SEASolEPI F PAW b hh} corresponding to $\eta=0.1$ goes higher and thinner at the increase of $\beta$ from $\beta=0.1$ to $\beta=0.5$. It shows that amplitude increases and the width decreases.

At the larger polarization $\eta=0.2$ described by blue (dashed) lines in Fig. \ref{SEASolEPI F PAW b hh} soliton shows different behavior. The increase of $\beta$ from $\beta=0.1$ to $\beta=0.5$ decreases the amplitude and width and this change is rather significant.

\section{CONCLUSIONS}

Earlier, analysis of the SSE in quantum plasmas allowed to predict the existence of SEAWs in degenerate e-i and e-p-i plasmas.
It also revealed existence of the SEPAWs in degenerate e-p-i plasmas. The spin-electron acoustic solitons propagating parallel to the external magnetic field was predicted as well. In spite the number of prediction of new phenomena, in this paper, we have demonstrated existence of spin-electron-positron acoustic solitons propagating parallel to the external magnetic field in degenerate e-p-i plasmas. We have made this prediction along with description of spin-electron acoustic solitons in e-p-i plasmas which were described earlier for e-i plasmas. We have shown influence of the spin polarization on the properties of positron-acoustic solitons which were well known for spinless e-p-i plasmas.

Altogether, we see that at weakly non-linear evolution e-p-i plasmas demonstrate existence of three solitons. Each of them is associated with one of three longitudinal waves having linear spectrum in the long-wavelength regime: SEAW, PAW, and SEPAW.

\section{\label{sec:level1}Acknowledgements}

The work of P.A. was supported by the Russian
Foundation for Basic Research (grant no. 16-32-00886) and the Dynasty foundation. The work of Z.I. was supported by the higher education commission Pakistan for the financial support under IRSIP Award No. 1-8/HEC/HRD/2015/4027.


\begin{thebibliography}{99}
\bibitem{Miller-1978} H. R. Miller and P. J. Witta, Active Galactic Nuclei,
Berlin Springer, p. 202 (1987).

\bibitem{Paul-astro-phys-2007} P. N. Arendt, Jr., and A. J. Eilek, Astrophys.
J. \textbf{581}, 469 (2002).


\bibitem{sturrock-Astrophys-J-164-1971} P. A. Sturrock, Astrophys. J. \textbf{164}, 529 (1971).

\bibitem{Han-Phys-Rev-D 86-2012} W. B. Han, R. Ruffini and S. S. Xue, Phys.
Rev. D \textbf{86}, 084004 (2012).

\bibitem{Iwamoto-Rev-E-47} N. Iwamoto, Phys. Rev. E \textbf{47}, 1 (1993).

\bibitem{zank-Phys-Rev-E-51-1995} G. P. Zank and R. G. Greaves, Phys. Rev.
E \textbf{51}, 9 (1995).

\bibitem{Wang -phys-plsma-20-2013} Y. Wang, P. K. Shukla and B. Eliasson,
Phys. Plasmas \textbf{20}, 013103 (2013).


\bibitem{Lai-Rev. Mod. Phys. 73-2001} D. Lai, Rev. Mod. Phys. \textbf{73}
629 (2001); A. K. Harding and D. Lai, Rep. Prog. Phys. \textbf{69}, 2631
(2006).

\bibitem{Shapiro-Germany-2004} S. L. Shapiro and S. A. Teukolsky, Black
Holes, White Dwarfs and Neutron Stars: The Physics of Compact Objects,
WILEY-VCH Verlag GmbH \& Co. KGaA, Weinheim, Germany (2004).



\bibitem{Davidson 1972} R. C. Davidson, Methods in Nonlinear Plasma Theory
(New York: Academic) p. 15 (1972).

\bibitem{Nicolson 1983} D. R. Nicolson, Introduction to Plasma Theory (New
York: Wiley) p. 171 (1983).

\bibitem{Hasegawa} A. Hasegawa, Plasma Instabilities and Nonlinear Effects
(New York: Springer) p. 34 (1975).


\bibitem{Pope-1975} S. I. Popel, S. V. Vladimirov and P. K. Shukla, Phys.
Plasmas \textbf{2}, 716 (1995).

\bibitem{Mahmood-new-j plasma-2003} S. Mahmood, A. Mushtaq, and H. Saleem,
New J. Phys. \textbf{5}, 28 (2003).


\bibitem{Haas-2003-pop} F. Haas, L. G. Garcia, J. Goedert, and G. Manfredi,
Phys. Plasmas \textbf{10}, 3858 (2003).



\bibitem{Azhar-2011} N. L. Tsintsadze, L. N. Tsintsadze, A. Hussain and G.
Murtaza, Eur. Phys. J. D \textbf{64}, 447 (2011).

\bibitem{Tsintsadze-j.plasma phy-2013} N. L. Tsintsadze, R. Chaudhary, A.
Rasheed, J. Plasma Phys. \textbf{79}, 587 (2013).

\bibitem{Nejoh--Aust- J-Phys} Y. N. Nejoh, Aust. J. Phys. \textbf{49}, 967
(1996).

\bibitem{Metref-pop-2014} H. Metref and M. Tribeche, Phys. Plasmas \textbf{21}, 122117 (2014).

\bibitem{Tribeche-pop-2009} M. Tribeche, K. Aoutou, S. Younsi and R. Amour,
Phys. Plasmas \textbf{16}, 072103 (2009).

\bibitem{Vagin-2006} D. V. Vagin, N. E. Kim, P. A. Polyakov and A. E.
Rusakov, Izv. RAN Ser. Fiz. \textbf{70} (3), 443 (2006) [Bull. Rus. Acad.
Sci.: Phys. \textbf{70} (3), 505 (2006)].

\bibitem{Moscow University Physics Bulletin-2007} P. A. Andreev and L.S.
Kuz'menkov, Moscow Univ. Phys. Bull. \textbf{62}, 271 (2007).

\bibitem{Browdn-PRL 101- (2008)} G. Brodin, M. Marklund, J. Zamanian, A.
Ericsson and P. L. Mana, Phys. Rev. Lett. \textbf{101}, 245002 (2008).

\bibitem{Misra-j-plasma phys-2010} A. P. Misra, G. Brodin, M. Marklund and
P.K. Shukla, J. Plasma Phys. \textbf{76}, 857 (2010).

\bibitem{Hussain PP 14 spin bernst} A. Hussain, M. Stefan and G. Brodin,
Phys. Plasmas \textbf{21}, 032104 (2014).

\bibitem{Andreev PRE 15} P. A. Andreev, Phys. Rev. E \textbf{91}, 033111 (2015).

\bibitem{Pavel-Zafar-2016} P. A. Andreev and Z. Iqbal, arXiv:1601.00761.

\bibitem{kuzmenkov- page-136-2001} L. S. Kuz'menkov, S. G. Maksimov, and V.
V. Fedoseev, Theor. Math. Phys. \textbf{126}, 110 (2001).

\bibitem{kuzmenkov- page-258-2001} L. S. Kuz'menkov, S. G. Maksimov, and V.
V. Fedoseev, Theor. Math. Phys. \textbf{126}, 212 (2001).

\bibitem{Pavel-soliton-pop-2016} P. A. Andreev, Phys. Plasmas \textbf{23},
012106 (2016).

\bibitem{Pavel- annal phys- 15} P. A. Andreev, Annals of Physics \textbf{361}, 278 (2015).

\bibitem{PA-arXiv-1409-2014 Landau damping} P. A. Andreev, arXiv:1409.7885.

\bibitem{Pavel-EPL-113 -2016} P. A. Andreev and L. S. Kuz'menkov, EPL
\textbf{113}, 17001 (2016).

\bibitem{Pavel-surface -2015} P. A. Andreev and L. S. Kuz'menkov,
arXiv:1512.07940.


\bibitem{P. A-superconductivity- Oct-2015} P. A. Andreev, P. A. Polyakov and
L. S. Kuz'menkov, arXiv:1507.03295.



\bibitem{Andreev AoP 14 exchange} P. A. Andreev, Annals of Physics \textbf{350}, 198 (2014).


\bibitem{Washimi-1966} H. Washimi and T. Taniuti, Phys. Rev. Lett. \textbf{17}, 996 (1966).

\bibitem{Svensson Astr J 82} R. Svensson, Astrophys. J. \textbf{258}, 335 (1982).

\bibitem{Akbari-Moghanjoughi PoP 10} M. Akbari-Moghanjoughi, Phys. Plasmas \textbf{17}, 082315 (2010).


\end{thebibliography}
\end{document}